\documentclass[10pt, conference, letterpaper]{IEEEtran}
\usepackage{cite}
\usepackage{graphicx}
\usepackage{psfrag}
\usepackage[caption=false,font=footnotesize]{subfig}
\usepackage{diagbox}
\usepackage{color}

\usepackage{url}
\usepackage{amsmath}
\usepackage[dvipsnames]{xcolor}
\usepackage{array}
\usepackage{graphics}
\usepackage{epsfig}
\usepackage{mathtools}
\usepackage{amsbsy}
\usepackage{amssymb}
\usepackage{amsthm}
\newtheorem{theorem}{Theorem}

\newtheorem{lemma}{Lemma}

\newtheorem{remark}{Remark}
\usepackage{multirow}
\newtheorem{definition}{Definition}

\usepackage{framed}
\usepackage{algorithm}
\newtheorem{proposition}{Proposition} 
\IEEEoverridecommandlockouts

\def\BibTeX{{\rm B\kern-.05em{\sc i\kern-.025em b}\kern-.08em
    T\kern-.1667em\lower.7ex\hbox{E}\kern-.125emX}}

\IEEEoverridecommandlockouts

\begin{document}

\title{Bandwidth Allocation and Service Differentiation\\ in D2D Wireless Networks}

\author{{Fran\c{c}ois Baccelli and Sanket S. Kalamkar}
\thanks{F. Baccelli (francois.baccelli@inria.fr) is with Inria-ENS, Paris, France and The University of Texas, Austin, USA. Sanket S. Kalamkar (sanket.kalamkar@inria.fr) is with Inria-LINCS Paris, France.}
\thanks{This work has received funding from the European Research Council (ERC) under the European Union's Horizon 2020 research and innovation programme grant agreement number 788851.}
}

\maketitle

\begin{abstract}

Inspired by a new feature in 5G NR called bandwidth part (BWP), this paper presents a bandwidth allocation (BA) model that allows one to adapt the bandwidth allocated to users depending on their data rate needs. Specifically, in adaptive BA, a wide bandwidth is divided into chunks of smaller bandwidths and the number of bandwidth chunks allocated to a user depends on its needs or type. Although BWP in 5G NR mandates allocation of a set of \textit{contiguous} bandwidth chunks, our BA model also allows other assumptions on chunk allocation such as the allocation of any set of bandwidth chunks, as in, e.g., LTE resource allocation, where chunks are selected uniformly at random. The BA model studied here is probabilistic in that the user locations are assumed to form a realization of a Poisson point process and each user decides independently to be of a certain type with some probability. This model allows one to quantify spectrum sharing and service differentiation in this context, namely to predict what performance a user gets depending on its type as well as the overall performance. This is based on exact representations of key performance metrics for each user type, namely its success probability, the meta distribution of its signal-to-interference ratio, and its Shannon throughput. We show that, surprisingly, the higher traffic variability stemming from adaptive BA is beneficial: when comparing two networks using adaptive BA and having the same mean signal and the same mean interference powers, the network with higher traffic variability performs better for all these performance metrics. With respect to Shannon throughput, we observe that our BA model is roughly egalitarian per Hertz and leads to a linear service differentiation in aggregated throughput value.


\end{abstract}

\section{Introduction}

\subsection{Motivation}
\label{sec:bwp_intro}
Our adaptive bandwidth allocation (BA) model is motivated by a new feature in 5G NR called bandwidth part (BWP)~\cite[Section 6.10]{NR_BWP_2019_1},
\cite[Section 4.4.5]{NR_BWP_2019_2}. The BWP concept is based on the division of a wide bandwidth into multiple \textit{contiguous} smaller chunks of bandwidth. The main aim of this division is to let the number of bandwidth chunks used by a wireless user depend on its type at a given time, namely on its current needs in terms of data rate or constraints in terms of hardware complexity and power consumption. This flexibility on what is allocated to users makes of BWP a new dimension of radio spectrum sharing.

\begin{table}[]
\caption{Types of wireless devices, wireless applications, and heterogeneous throughput demands. The numbers in parentheses refer to 2017 and 2022, respectively. (Sources: Cisco VNI Mobile, 2019~\cite{Cisco_VNI} and MediaTek white paper~\cite{mediatek}.)}
\begin{tabular}{|l|l|l|l|l|}
\hline
\multirow{3}{*}{} & \multicolumn{2}{l|}{\!\!\!\! Wireless devices~\cite{Cisco_VNI}} & \multicolumn{2}{l|}{\!\!\! Wireless applications~\cite{mediatek}\!\!\!} \\ \cline{2-5} 
 & \multirow{2}{*}{\begin{tabular}[c]{@{}l@{}}\!\!\! \% Share \\ \!\!\!(2017, 2022)\!\!\!\end{tabular}} & \multirow{2}{*}{\begin{tabular}[c]{@{}l@{}}\!\!\! \% Growth \!\!\!\\ \!\!\!(2017, 2022)\!\!\!\!\!\end{tabular}} & \multicolumn{2}{l|}{\multirow{2}{*}{\!\!\!\!\! Throughput demand (Mbps)}\!\!\!} \\
 &  &  & \multicolumn{2}{l|}{} \\ \hline
\multirow{2}{*}{\begin{tabular}[c]{@{}l@{}}\!\!\! \!Smartphones \\ + phablets\!\!\end{tabular}} & \multirow{2}{*}{\!\!\!(50, 44)} & \multirow{2}{*}{\!\!\!(88, 93)} & \multirow{2}{*}{\begin{tabular}[c]{@{}l@{}}\!\!\! \!\!Streaming\!\!\!\!\!\!\!\!\! \\ \!\!\!\!\! video\end{tabular}} & \!\!\!Video (1.5) \\ \cline{5-5} 
 &  &  &  & \!\!\!HD video (5) \\ \hline
\multirow{2}{*}{\!\!\!M2M} & \multirow{2}{*}{\!\!\!(11, 31)} & \multirow{2}{*}{\!\!\!(1.8, 2.2)} & \multirow{2}{*}{\begin{tabular}[c]{@{}l@{}}\!\!\!\!\! Online \\ \!\!\!\!\! gaming\end{tabular}} & \!\!\!Min (1) \\ \cline{5-5} 
 &  &  &  & \!\!\!Full (25) \\ \hline
\multirow{3}{*}{\!\!\!Nonsmartphones \!\!\!\!\!} & \multirow{3}{*}{\!\!\!(34, 10)} & \multirow{3}{*}{\!\!\!(1.3, 0.3)} & \multirow{3}{*}{\begin{tabular}[c]{@{}l@{}}\!\!\!\!\! Video \\\!\!\! \!\!\! service\\  \!\!\!\!\! \& sharing \!\!\!\end{tabular}} & \!\!\!Min (0.5) \\ \cline{5-5} 
 &  &  &  & \!\!\!HD video (2) \\ \cline{5-5} 
 &  &  &  & \begin{tabular}[c]{@{}l@{}}\!\!\!HD video \\ \!\!\!sharing\! (10)\end{tabular} \\ \hline
\multirow{3}{*}{\!\!\!Tablets} & \multirow{3}{*}{\!\!\!(2, 3)} & \multirow{3}{*}{\!\!\!(4.6, 2.9)} & \multirow{3}{*}{\begin{tabular}[c]{@{}l@{}}\!\!\!\!\! VoIP \end{tabular}} & \!\!\!Voice (0.1) \\ \cline{5-5} 
 &  &  &  & \!\!\!HD video (1.5)\!\!\!\!\! \\ \hline
\multirow{2}{*}{\!\!\!PCs} & \multirow{2}{*}{\!\!\!(2, 1)} & \multirow{2}{*}{\!\!\!(4.3, 1.6)} & \multirow{2}{*}{\begin{tabular}[c]{@{}l@{}}\!\!\!\!\! Social \\ \!\!\!\!\! media\end{tabular}} & \multirow{2}{*}{\!\!\!Text (0.3)} \\
 &  &  &  &  \\ \hline
\end{tabular}
\label{tab:device_types}
\end{table}

The BWP setting requires the allocation of a set of contiguous bandwidth chunks. But in general, e.g., in carrier aggregation in LTE, there is no restriction on a user to use contiguous bandwidth chunks~\cite{CA}. In this paper, we propose a general adaptive BA model that allows one to analyze both the non-contiguous bandwidth chunk allocation of LTE and the contiguous case of BWP.

Such a bandwidth adaptation depending on the user type is particularly important in future wireless networks, e.g., $5$G networks, which need to accommodate a larger variety of wireless devices (see Table~\ref{tab:device_types}), running in turn a larger variety of wireless applications with highly heterogeneous throughput demands (see Table~\ref{tab:device_types}). More specifically, as shown in Table~\ref{tab:device_types}, mobile video streaming constitutes the majority of wireless traffic and requires higher data rate and hence wider spectrum allocation
in the adaptive BA setting. In contrast, text and e-mail applications have lower data rate requirements and would thus be allocated less bandwidth. For users of the latter type, the use of wide bandwidth leads to high costs, in particular, high idling power consumption by radio-frequency (RF) and baseband signal processing circuitry. Hence, the use of different bandwidth sizes allows a balance between data rate variations and power consumed by users.

One typical use case is that of web browsing, where the user is active for a short time on wide bandwidth to accommodate the bursty traffic (download of a web page with pictures), and then stays active on narrow bandwidth for the time until it again encounters bursty traffic situation. This power-saving feature of adaptive BA helps making wide bandwidth operations energy more efficient.

The benefits of the flexibility offered by adaptive BA are hence multiple. Not only this adaptation saves energy for those applications and devices with lower data rate requirement or bursty traffic, but this in turn diminishes the interference incurred by other nearby devices of all types. Indeed, due to the broadcast nature of the wireless medium, users interact with each other through mutual interference. The wider the bandwidth, the higher the interference. Hence the bandwidth adaption  has two competing effects. On the positive side, a wider bandwidth increases the signal power and hence the throughput. On the negative side, it increases the interference power, which has a detrimental collective effect. As a result, it is fair to say that there is no global understanding of the effect of adaptive BA on the per-type and the overall performance. 

To the best of our knowledge, there is no analytical study of a resource allocation model that  captures the features of BWP. The existing studies on BWP (discussed in Section~\ref{sec:rel_work}) are limited to the investigation of power savings and bandwidth switching using simulations. The study of the key wireless network performance metrics is an open area. In particular, there is currently no known way to predict the effect of the BA on the performance of a user of a given type.

The main motivation of the present paper can now be stated in simple terms: it is to provide a statistical model allowing one to analyze adaptive bandwidth allocation motivated by BWP in device-to-device (D2D) wireless networks, and more precisely to predict the key performance metrics of the typical user of a given type in this context. It is appropriate to stress the analogy with the theory of differentiated services in wireline networks (DiffServ~\cite{DiffServ}), which was instrumental in classifying and managing different types (classes) of network traffic and in predicting their interactions. The aim of this paper is to make a first step in the direction of a quantitative theory for BWP-based bandwidth allocation and the management of different types of network traffic in this class of wireless networks.

\subsection{Contributions}
\textbf{An adaptive BA model.} The first contribution of this paper is a stochastic model for adaptive BA motivated by BWP in infrastructureless wireless networks,\footnote{In 5G NR, although the BWP feature is proposed for cellular networks, we focus on an infrastructureless network (a simpler one compared to the cellular network) since this paper is the first attempt to analytically study a bandwidth allocation scheme motivated by BWP.}  where the type of the user is determined by the number of bandwidth chunks it uses which in turn adapts to the current user needs. As already explained, our model accommodates the following two BA approaches.
\begin{itemize}
\item Contiguous BA: A user that needs $k$ bandwidth chunks can select any set of $k$ \textit{contiguous} chunks uniformly at random. Such a contiguous bandwidth allocation may be used in the BWP setting in 5G NR.
\item Random BA: A user that needs $k$ bandwidth chunks can select any set of $k$ chunks uniformly at random. Such a bandwidth allocation has applications in LTE resource allocation, e.g., carrier aggregation~\cite{CA}.
\end{itemize} 
Also, our adaptive BA model allows us to quantify service differentiation in wireless networks by capturing interaction among different types of wireless users. Specifically, the model allows one to calculate the performance achieved by each type of user.

\textbf{Performance analysis.} Using tools from stochastic geometry for D2D networks, we derive analytical expressions for key wireless network performance metrics, namely, success probability, meta distribution of the signal-to-interference ratio (SIR), Shannon throughput, and Shannon throughput per Joule. These expressions permit the evaluation of per-type and overall performance metrics.

\textbf{Different performance viewpoints.} Our model allows one to analyze the bandwidth allocation from the viewpoints of both users and operators. Per-type performance, although relevant to operators, is more important for users, while overall performance might be important too from the operator's viewpoint due to, e.g., the link between this and pricing. Thus, the machinery proposed here to predict both could be useful to help an operator make a choice between the following options: 1) allocate the entire bandwidth, 2) adaptively allocate bandwidth chunks.  

\textbf{The mean model.} The proposed BA model introduces additional randomness due to the probabilistic selection of the set of bandwidth chunks depending on the user type. We show that the increased variability in traffic due to adaptive BA may improve the performance for the same mean interference and the same mean signal powers. This is particularly useful when comparing two networks based on adaptive BA but with different mix of user types. 

\textbf{Service differentiation.} We show that adaptive BA leads to a roughly egalitarian service in Shannon throughput per Hz and to a linear service differentiation in aggregated Shannon throughput.

\subsection{Related work}
\label{sec:rel_work}
In wireline networks, traffic-flow characterization has received significant attention for networks consisting of users running different types of applications. To understand the behavior of heterogeneous flows and their interactions, flows have been classified based on their features, e.g., traffic size (as \textit{elephant} and \textit{mouse})~\cite{traffic_size_Thompson, traffic_size_Papagiannaki, traffic_size_Estan}.
For wireless networks as well, in the adaptive BA setting, one can make an analogy to elephants and mice: users needing wide bandwidth can be viewed as \textit{elephants}, while users needing small bandwidth as \textit{mice}.


As alluded to earlier, the $3$GPP has very recently considered the inclusion of BWP in $5$G NR to enable spectrum flexibility and power savings. The literature on how BWP affects power savings, throughput, and reliability is very limited. For instance, \cite{Jeon_BWP_2018} discusses power savings due to BWP. Since a user need not transmit or receive outside the bandwidth allocated to it, the user consumes less power in some scenarios, for example, involving bursty traffic. The work in~\cite{Fuad_BWP} shows that the bandwidth switching to save power results in increased latency and decreased throughput for low load and bursty traffic. The work in~\cite{Arslan_reliability} studies the effect of BWP on reliability and fairness in wireless networks. But these works on BWP use simulations as a tool to evaluate the performance based on the dynamic BWP management. 

This paper is focused on adaptive BA for D2D networks. There exists a large number of works on spectrum allocation including multi-channel scenarios in standalone D2D networks or D2D networks sharing spectrum with cellular networks. For instance, see~\cite{Stefanatos_BWP,Kyasanur_BWP,Sun_BWP} and references therein. In relation to heterogeneity among devices in a D2D network based on allocated bandwidth, \cite{Yuan_BWP} focuses on the dynamic allocation of bandwidth to unlicensed users based on data rate demand, provided the allocated spectrum is unoccupied by licensed users and other unlicensed users. This results in an orthogonal bandwidth allocation to avoid mutual interference. Using a simulation approach, the work in \cite{Hamidouche_BWP} takes battery life into account and tries to maximize the number of completed transmissions as a function of already allocated bandwidth. 

Stochastic geometry, which is the main tool used in this paper, has been extensively used to model and analyze both infrastructureless (e.g., D2D) and infrastructure (e.g., cellular) networks~\cite{bb_book,mh_book,SG_tut,SG_cellular}. Especially, Poisson point process (PPP)-based models are very popular for the analysis of wireless networks. Heterogeneous PPP based cellular networks were in particular discussed in \cite{SG_het}. These models feature several types of base stations and a single class of users associating e.g. to the closest base station. In contrast, the setting discussed here features different types of transmitters with dedicated receivers and adapting their bandwidths to their needs.

In \cite{Jindal_BWP} the authors aim to maximize the density of successful transmissions given an outage constraint at the typical user. The user always selects one bandwidth chunk uniformly at random irrespective of its need. For the same bandwidth partitioning scheme as~\cite{Jindal_BWP}, \cite{Haenggi_BWP_2014} analyzes the mean local delay and \cite{sanket_globecom} exhibits the tradeoff between the density of successful transmissions and the mean local delay. As in~\cite{Jindal_BWP}, the work in~\cite{Lu_BWP} maximizes the density of successful transmissions, but for frequency-selective channels, where again only one bandwidth chunk is selected for the transmission. To the best of our knowledge, no concrete adaptive BA models capturing BWP features and accommodating different types of users are available in the literature.

\section{System Model}
\label{sec:models}

\subsection{Network model}
We consider infrastructureless wireless networks such as ad hoc, D2D, and machine-to-machine (M2M) networks.\footnote{As shown in Table~\ref{tab:device_types}, in year 2022, $31\%$ of mobile devices  are expected to belong to M2M networks.} The transmitters are randomly located according to a homogeneous PPP $\Phi \subset \mathbb{R}^2$ of intensity $\lambda$. Each transmitter has a receiver at fixed distance $R$ in a random direction~\cite{bb_book}. Since the homogeneous PPP is stationary, one can just focus on the reference link between a receiver at the origin $o$ and its associated transmitter at $x_0 \in \Phi$ with $\|x_0\| = R$. Averaging over $\Phi$, this representative link becomes the {\em typical} link in that it has the same statistical properties as those obtained by averaging over all other links in the network.

A transmission is subject to some path loss, where the path-loss function is given by $\ell(r)$ for distance $r$. Furthermore, the transmissions experience Rayleigh fading, where the channel power gain is an exponential random variable with mean $1$. Let $h_x$ denote the channel power gain between the typical receiver at the origin and the transmitter at $x \in \Phi$ due to fading. Then $h_x \sim \exp(1)$. We focus on the interference-limited scenario, where the noise power is negligible compared to the interference power.
\subsection{Bandwidth allocation model}
\label{sec:bwp_model}
In this subsection, we describe our two BA models.

Let $W$ be the total bandwidth available to users. Without loss of generality, we assume $W = 1$. 

\textbf{Random BA model.} In the case without continuity requirement, our adaptive BA model is as follows:
\begin{itemize}
\item[(1)] The total bandwidth is divided into $K$ orthogonal chunks of equal bandwidth of $1/K$.
\item[(2)] Depending on how many chunks a transmitter uses, the transmitters are categorized into $K$ types. A type-$i$ transmitter selects $i$ ($ 1\leq i \leq K$) chunks for its transmission. In other words, the type of the user is decided by the amount of bandwidth used by that user. Let $\mathcal{T} = \lbrace \mathcal{T}_1, \mathcal{T}_2, \dotsc, \mathcal{T}_K \rbrace$, where $\mathcal{T}_i$ is the set of all subsets $\mathcal{T}_{i,q}$ of $[K] \triangleq \lbrace 1, 2, \dotsc, K\rbrace$ of cardinality $i$, i.e., $|\mathcal{T}_{i,q}| = i$ with $q = 1, 2, \dotsc, {K \choose i}$. Here, ${K \choose i}$ is the number of possible ways of selecting $i$ chunks from $K$ chunks, i.e., $|T_i| = {K \choose i}$. For example, for $K = 3$, a type-$2$ transmitter, i.e., $i = 2$, can select two from three chunks $1$, $2$, and $3$, i.e., we have $\mathcal{T}_2 = \lbrace \lbrace 1, 2\rbrace, \lbrace 2, 3\rbrace, \lbrace1, 3\rbrace\rbrace$.
\item[(3)] A type-$i$ transmitter further selects a set of chunks from $\mathcal{T}_i$ uniformly at random. For the aforementioned example with $K = 3$, a type-$2$ transmitter selects $2$ chunks for transmission, and it does so by selecting one of the possible sets of chunks from $\lbrace 1, 2\rbrace, \lbrace 2, 3\rbrace$, and $\lbrace 1, 3\rbrace$ at random.
\item[(4)] Each transmitter independently decides to be of $i$th type with probability $p_i$ with $\sum_{i = 1}^{K} p_i = 1$.
\end{itemize}

\textbf{Contiguous BA model.} The case of contiguous bandwidth chunks is a variant of the above random BA model because the set of contiguous chunks is a subset of $\mathcal{T}_i$. For a type-$i$ user, there are $K-i+1$ sets of contiguous chunks. Again, a natural way for type-$i$ user to select one of $K-i+1$ sets of contiguous chunks is to do so uniformly at random.\footnote{The contiguous BA model is also random in nature but with a restriction that only contiguous bandwidth chunks are allocated.}  For example, for $K = 3$ and $i = 2$, we have $\mathcal{T}_2 = \lbrace \lbrace 1, 2\rbrace, \lbrace 2, 3\rbrace\rbrace$, and the type-$2$ transmitter can select one set from $\lbrace 1, 2\rbrace$ and  $\lbrace 2, 3\rbrace$ uniformly at random.

\textbf{User types.} For both BA models, the probabilities $p_i$, $1 \leq i \leq K$, quantify service requirements. Specifically, these probabilities can be obtained from the statistical analysis of the user traffic. For instance, $p_i$ could be set according to what proportion of user's data traffic consists of large data transfer such as video streaming and what proportion consists of small traffic such as text or e-mail transmissions. As Table~\ref{tab:device_types} in Section~\ref{sec:bwp_intro} shows, a user's traffic consists of traffic from different wireless applications with heterogeneous throughput demands. Thus, the probabilities $p_i$ could be set according to the user's throughput demands. Let us consider a use case: \cite{Cisco_VNI} reports that, in 2018 approximately 63\% of the traffic of a mobile user was video content, and it will grow to 79\% in 2022. Due to higher throughput demand for video, a mobile user is likely to request a wider bandwidth (and hence a large number of bandwidth chunks).  



\subsection{Signal-to-interference ratio (SIR)}
\label{sec:sir}
In interference-limited wireless networks, many key performance metrics are based on SIR. The SIR at the typical receiver located at the origin $o$ with respect to its associated transmitter at $x_0 \in \Phi$ is given by $\mathsf{SIR}_o \triangleq  \frac{S}{I}$, where $S$ and $I$ are the received signal power and the interference power at the origin, respectively. 

The received SIR depends on the type of the typical user because the signal and interference powers are functions of the bandwidth allocated to the user. Without loss of generality, we condition on the fact that the typical transmitter is of type $k$, i.e., it uses $1 \leq k \leq K$ chunks for transmission. 

\textbf{Signal power}: We assume that a transmitter spends power $P$ per chunk used for a transmission.\footnote{This assumption is in line with the proposed BWP model for $5$G, where the transmit power is mentioned in terms of the power spectral density.} Hence, $P$ is expressed in Joule-s$^{-1}$-Hz$^{-1}$. The received signal power at the typical receiver is given by $S_k = kPh_{x_0}\ell(x_0)$, since the typical transmitter selects $k$ chunks for transmission. Here, $h_{x_0}\sim\exp(1)$ denotes the channel power gain on the typical link.

\textbf{Interference power}: 
 Let $t_x^{(k)}$ denote the number of such overlapping  chunks between an interferer at $x \in \Phi$ and the typical transmitter (of type $k$). Then, the interference power at the typical receiver is given by
\begin{align}
I_k = \sum_{x \in \Phi\setminus\lbrace x_0 \rbrace} P t_x^{(k)} h_x  \ell(x).
\label{eq:intf_pow_main}
\end{align} 
Our assumption in this paper is that a type-$i$ interfering transmitter selects $i$ chunks with probability $p_i$ independently of other transmitters. Hence, the original PPP $\Phi\setminus\lbrace x_0\rbrace$ of interferers can be split into $K$ independent PPPs of intensities $\lambda p_i$, $i = 1, 2, \dotsc, K$. Let $\Phi_{i} \setminus \lbrace x_0\rbrace$ denote the PPP of type-$i$ interferers. In random BA, the typical transmitter selects a set $T_{k, u}$ of $k$ chunks at random with $1 \leq u \leq {K \choose k}$. In contiguous BA, the typical transmitter selects a set $T_{k, u}$ of $k$ contiguous chunks at random with $1 \leq u \leq K-k+1$. Also, in random BA, a type-$i$ interferer selects a set $T_{i,v}$  of $i$ chunks with $1 \leq v \leq {K \choose i}$ at random and independently. Similarly, in contiguous BA, a type-$i$ interferer selects a set $T_{i,v}$ of $i$ contiguous chunks with $1 \leq v \leq K-i+1$ at random and independently. Then the interference power from the transmitter at $x \in \Phi\setminus \lbrace x_0\rbrace$ to the typical receiver is $P|T_{k, u} \cap T_{i, v}|h_x\ell(x)$, where $|T_{k, u} \cap T_{i, v}|$ is the number of overlapping chunks between a type-$i$ interferer and the typical transmitter. Let $0 \vee (i+k-K) \leq t \leq k \wedge i \leq K$ denote the number of overlapping chunks between a type-$i$ interfering transmitter and the typical transmitter, where $0 \vee (i+k-K)$ means $\max(0, i+k-K)$ and $k \wedge i$ means $\min(k, i)$. Note that $t$ is a random variable since the typical and interfering transmitters select $k$ and $i$ chunks, respectively, uniformly at random. For notation simplicity, we do not always indicate the dependence of $t$ on $k$ and $i$. 

Depending on the number of overlapping chunks $t$, the PPP $\Phi_i$ can further be partitioned into $t$ independent PPPs denoted by $\Phi_{i, t}$. The PPP $\Phi_{i, t}$ corresponds to transmitters located at $x \in \Phi_i$ that have $0 \vee (i+k-K) \leq t \leq k \wedge i$ chunks in common with the typical transmitter. Consequently, the interference power received at the typical receiver at the origin from type-$i$ interferers is given by
\begin{align}
I_{k,i} =  \sum_{t = 0 \vee (i+k-K)}^{k \wedge i}\sum_{x \in \Phi_{i,t}\setminus \lbrace x_0 \rbrace} tPh_x\ell(x).
\label{eq:intf_pow_i}
\end{align}
The total interference power at the typical receiver of type $k$ follows as $I_k = \sum_{i = 1}^{K} I_{k,i}$.

\textbf{SIR expression}: Following the expressions of signal and interference powers, the SIR experienced at the typical user of type $k$ can be expressed as
\begin{align}
\mathsf{SIR}_o^{(k)} = \frac{kh_{x_0}\ell(x_0)}{\sum_{i = 1}^{K} \sum_{t = 0 \vee (i+k-K)}^{k \wedge i}\sum_{x \in {\Phi_{i,t}}\setminus \lbrace x_0 \rbrace} th_x\ell(x)}.
\label{eq:sir_main_ex}
\end{align}
The transmit power $P$ per chunk vanishes from \eqref{eq:sir_main_ex}.

\section{Performance Metrics}
In this section, we define and discuss the three performance metrics that we consider in this work. These metrics are based on the SIR received at the typical receiver and shed light on different aspects of the performance of users. 

Although the definitions of the performance metrics are general, the calculation of these metrics depends on the type of the typical user. The overall performance can be evaluated by unconditioning with respect to the type of the typical transmitter. For instance, let $f_k$ be some performance function conditioning on the fact that the typical transmitter is of type $k$. Then, the overall performance $f$ of the typical user is given by $f = \sum_{k = 1}^{K} p_k f_k$. For the notation simplicity, in this section, we do not explicitly show the dependence of the performance metrics on $k$ while defining them. However, we shall reintroduce the parameter `$k$' in the performance analysis done in Section~\ref{sec:perf_ana}. 

\subsection{Success probability}
\begin{definition}[Success probability]
The success probability of the typical user at the origin $o$ is the complementary cumulative distribution function (ccdf) of the SIR, which is
\begin{align}
p_{\rm s}(\theta) \triangleq \mathbb{P}^{!}_{o}(\mathsf{SIR}_o > \theta),
\label{eq:p_suc}
\end{align}
where $\theta \in \mathbb{R}^{+}$ is the target SIR threshold. 
\end{definition}
\noindent Here, $\mathbb{P}^{!}_{o}(\cdot)$ denotes the reduced Palm probability of the receiver point process. $p_{\rm s}$ is an outage-based performance metric: if the received SIR at the typical receiver is larger than $\theta$, the transmission is considered successful. 

When the underlying point process is ergodic, $p_{\rm s}$ can also be interpreted as the fraction of concurrent transmissions that achieve an SIR greater than $\theta$ in each realization of the network. In other words, $p_{\rm s}$ is nothing but a {\em spatial average} in that it is evaluated by taking a certain expectation over the point process. This average is certainly very useful in wireless networks, but it does not provide information about individual user success probabilities. 
Hence, to analyze the fine-grained performance, we need to quantify how individual user success probabilities are distributed around the average $p_{\rm s}$. The meta distribution of the SIR defined below is one such fine-grained performance metric in wireless networks~\cite{Haenggi_MD_2016}.

\subsection{Meta distribution of the SIR}
\label{sec:md_sir}
We are interested in the random variable $P_{\rm s}$ defined as $P_{\rm s}(\theta, \Phi) \triangleq \mathbb{P}(\mathsf{SIR} > \theta \mid  \Phi)$, where the conditional probability is taken over the fading and the random channel access scheme of interferers determined by the BA model. This conditional random variable is the probability that the fading and the random channel access scheme yielding an SIR at least $\theta$ for the user under consideration for a given realization of $\Phi$. Hence, $P_{\rm s}$ is the success probability of that user conditioned on the point process $\Phi$. The distribution of  $P_{\rm s}$ obtained by taking an expectation over the point process is the meta distribution of the SIR. Formally, it is defined as follows.
\begin{definition}[Meta distribution]
The meta distribution of the SIR is the distribution function
\begin{align}
\bar{F}(\theta, x) \triangleq \mathbb{P}^{!}_{o}(P_{\rm s}(\theta, \Phi) > x), \quad \theta \in \mathbb{R}^{+}, x \in [0, 1].
\label{eq:md_def}
\end{align}
\end{definition}

\noindent The meta distribution $\bar{F}(\theta, x)$ is the probability that the user under consideration has a reliability at least $x$ for the target SIR threshold of $\theta$, where the reliability  is the conditional success probability $P_{\rm s}(\theta, \Phi)$. When the underlying point process is ergodic (such as the PPP), $\bar{F}(\theta, x)$ can be interpreted as the fraction of users that achieve the target SIR of $\theta$ with probability at least $x$. The parameter $x$ can be viewed as the target reliability. Note that the standard success probability $p_{\rm s}(\theta)$ given in \eqref{eq:p_suc} is the mean of the conditional random variable $P_{\rm s}(\theta, \Phi)$. Hence, the meta distribution of the SIR $\bar{F}(\theta, x)$ provides much sharper SIR performance compared to its mean $p_{\rm s}(\theta)$.
%

\subsection{Shannon throughput}
The success probability $p_{\rm s}(\theta)$ and the meta distribution $\bar{F}(\theta, x)$ correspond to the binary event whether the SIR is larger than some threshold $\theta$ or not. Hence, they fail to use the SIR values larger than $\theta$ and reduce the SIR threshold $\theta$ to avoid outages if the SIR value is smaller than $\theta$. Instead, a transmitter can (if possible) adapt to channel conditions and adjust the SIR threshold $\theta$ to the maximum value such that $\mathsf{SIR} \geq \theta$ (alternatively, adapt the coding rate). In this case, Shannon throughput is a more suitable performance metric to quantify the performance in wireless networks.
\begin{definition}[Shannon throughput]
For the bandwidth $B$ used by a user, the Shannon throughput is
\begin{align}
\mathcal{R} \triangleq B\mathbb{E}(\log_2(1+\mathsf{SIR})).
\label{eq:shannon_throughput}
\end{align} 
\end{definition}
\noindent The Shannon throughput is the average of the instantaneous throughput $B\log_2(1+\mathsf{SIR})$ of a random user in the network with adaptive coding, where the SIR corresponds to that random user. The Shannon throughput is expressed in bits/s.

\section{Performance Analysis}
\label{sec:perf_ana}
In this section, we calculate the expressions of the performance metrics.

\subsection{Success probability}
We obtain a closed-form expression of the success probability. Although our analysis can be generalized to arbitrary path loss models, we first focus on the standard power-law path loss function given as $\ell(x) = \|x\|^{-\alpha}$ with $\alpha > 2$ being the path loss exponent. We later provide a simple closed-form expression of the success probability for the bounded path loss function $\ell(x) = (c_0 +\|x\|^\alpha)^{-1}$ with $c_0 > 0$.
\begin{lemma}
\label{lem:suc_prob_k}
Conditioned on the typical transmitter being of type $k$, i.e., when $x_0 \in \Phi_k$, the success probability $p_{\rm s}^{(k)}$ of the typical receiver at the origin can be expressed as
\begin{align}
p_{\rm s}^{(k)}(\theta) & = \prod_{i = 1}^{K} p_{\rm s} ^{(k,i)}(\theta),
\label{eq:cond_suc}
\end{align}
where $p_{\rm s} ^{(k,i)}$ is the success probability of the typical receiver due to the interference from type-$i$ interferers only conditioned on the typical transmitter being of type $k$.
\end{lemma}
\begin{IEEEproof} 
See Appendix~\ref{app:suc_prob_k}.
\end{IEEEproof}
The probabilities $p_{\rm s} ^{(k,i)}$ quantify the performance of differentiated services. 
Specifically, as shown in the following theorem, by just playing with the values of $k$ and $i$, one can investigate how elephants (users with wider bandwidth) affect the performance of mice (users with smaller bandwidth) and vice-versa. The following theorem gives a simple closed-form expression of $p_{\rm s} ^{(k,i)}$ that allows one to quantify the effect of type-$i$ users on the success probability of a type-$k$ user.
\begin{theorem}
\label{thm:suc_ki}
Let us condition on the typical transmitter being a type-$k$ user. For the power-law path loss model, the success probability $p_{\rm s} ^{(k,i)}$ of the typical user is
\begin{align}
p_{\rm s} ^{(k,i)}(\theta) = \exp\left(-\lambda p_i C \theta^{\delta} \sum_{t = 0 \vee (i+k-K)}^{k \wedge i}p^{(t)}_{k,i}\left(\frac{t}{k}\right)^\delta\right),
\end{align}
where $C \triangleq \pi R^2 \Gamma(1+\delta)\Gamma(1-\delta)$ with $\delta \triangleq 2/ \alpha$ and $p^{(t)}_{k,i}$ is the probability that an interferer of type $i$ has $0 \vee (i+k-K) \leq t \leq k\wedge i$ chunks in common with the typical transmitter, conditioned on the fact that the latter is a type-$k$ user. For the random BA case, we have
\begin{align}
p^{(t)}_{k,i} = \frac{{k \choose t}{K-k \choose i-t}}{{K \choose i}},
\label{eq:pt}
\end{align}
whereas for the contiguous BA case,

{{\small\begin{align}
p^{(t)}_{k,i} =\begin{cases}
\frac{2(K+t-k-i+1)}{(K-k+1)(K-i+1)} & \mathrm{if}~t < k \wedge i~\mathrm{and}~K \geq k+i-t\\
0 & \mathrm{if}~t < k \wedge i~\mathrm{and}~K < k+i-t\\
\frac{i-k+1}{K-k+1} & \mathrm{if}~t = k~\mathrm{and}~k \leq i \\
\frac{k-i+1}{K-i+1} & \mathrm{if}~t = i~\mathrm{and}~k > i \\
\frac{(K-k-i+1)(K-k-i+2)}{(K-k+1)(K-i+1)} & \mathrm{if}~t = 0~\mathrm{and}~K \geq k+i.
\end{cases} 
\label{eq:pt_cont}
\end{align}}}
\end{theorem}
\begin{IEEEproof}
See Appendix~\ref{app:suc_ki}.
\end{IEEEproof}
Finally, unconditioning on the type of the typical transmitter yields the overall success probability $p_{\rm s} = \sum_{k = 1}^{K}p_k p_{\rm s}^{(k)}(\theta)$ as
\begin{align}
p_{\rm s}(\theta) = \sum_{k = 1}^{K} p_k \exp\!\left(\!\!-\lambda C \theta^{\delta}\sum_{i = 1}^{K}p_i  \sum_{t = 0 \vee (i+k-K)}^{k \wedge i}p^{(t)}_{k,i} \left(\frac{t}{k}\right)^\delta\right).
\label{eq:int_suc} 
\end{align}

\textbf{Bounded path loss function}: For the bounded path loss function $\ell(x) = (c_0 + \|x\|^{\alpha})^{-1}$ with $c_0 > 0$, we have 
{{\small
\begin{align}
p_{\rm s}^{(k,i)}(\theta) = \exp\!\left(\!\!- p_i C_b\theta \!\!\!\!\!\!\!\sum_{t = 0 \vee (i+k-K)}^{k \wedge i}\!\!\!\!\!\!\!\!p^{(t)}_{k,i}\!\left(\frac{t}{k}\right)\!\! \!\left(\frac{\theta t}{k}(c_0 + R^{\alpha}) + c_0\!\right)^{\delta{-}1}\! \right)\!\!,
\end{align}}}\vspace*{-4mm}

\noindent where $C_b \triangleq \lambda \pi (c_0+R^{\alpha})\Gamma(1+\delta)\Gamma(1-\delta)$. The proof follows the one of Theorem~\ref{thm:suc_ki}. The standard path loss function $\ell(x) = \|x\|^{-\alpha} $ is a special case of the bounded path loss function $(c_0 + \|x\|^{\alpha})^{-1}$ with $c_0 = 0$. The equations \eqref{eq:cond_suc} and \eqref{eq:int_suc} are also valid for the bounded path loss case.

\subsection{Meta distribution of the SIR}

The direct calculation of the SIR meta distribution, i.e., the ccdf in \eqref{eq:md_def}, is impossible. Hence we take an indirect route, where we first calculate $b$th moments ($b \in \mathbb{C}$) of $P_{\rm s}$ and use those moments to obtain the ccdf $\bar{F}(\theta, x)$ accurately using the Gil-Pelaez theorem~\cite{Gil_Pelaez} or approximately by matching the first and second moments to those of the beta distribution~\cite{Haenggi_MD_2016}. In other words, averaging over the point process $\Phi$ and averaging over the fading and the random channel access scheme are done separately. This is unlike the calculation of the success probability $p_{\rm s}$, where averaging over $\Phi$, the fading, and the random channel access scheme are done simultaneously.

\begin{theorem}
\label{thm:Mb_k}
Conditioning on the typical transmitter being of type $k$, the $b$-th moment of $P_{\rm s}^{(k)}$ is
{{\small\begin{align*}
M_{b}^{(k)} = \exp\!\!\left(\!\!-2\pi\lambda \!\!\int_{0}^{\infty}\!\!\!\left[\!1-\!\!\left(\sum_{i = 1}^{K}p_i\!\!\!\!\sum_{t = 0 \vee (i+k-K)}^{k \wedge i}\!\frac{p^{(t)}_{k,i}}{1+\theta \frac{t}{k}\frac{\ell(r)}{\ell(x_0)}}\!\!\right)^{b}\right]\!\mathrm{d}r\!\!\right)\!\!,
\end{align*}}}
where $p^{(t)}_{k,i}$ is given by \eqref{eq:pt} for random BA and by \eqref{eq:pt_cont} for contiguous BA.
\end{theorem}
\begin{IEEEproof}
See Appendix~\ref{app:Mb_k}.
\end{IEEEproof}

\subsection{Shannon throughput}
\begin{theorem}
Conditioning on the typical transmitter being of type $k$, the Shannon throughput $\mathcal{R}^{(k)}$ is given by
\begin{align}
\mathcal{R}^{(k)} = \frac{k}{K} \int_{0}^{\infty} p_{\rm s}^{(k)}(2^y-1)~\mathrm{d}y,
\end{align}
where $p_{\rm s}^{(k)}$ is given by~\eqref{eq:cond_suc}.
\end{theorem}
\begin{IEEEproof}
A transmitter of type $k$ has $k/K$ Hz of bandwidth for the transmission. Hence, we have
\begin{align}
\mathcal{R}^{(k)} &= \frac{k}{K}\int_{0}^{\infty} \mathbb{P}(\log_2(1 + \mathsf{SIR}_o) > y \mid x_0 \in \Phi_k)~\mathrm{d}y \nonumber \\
&= \frac{k}{K} \int_{0}^{\infty} p_{\rm s}^{(k)}(2^y-1)~\mathrm{d}y.
\label{eq:Shannon_throughput_no_joule}
\end{align}
\end{IEEEproof}

The overall Shannon throughput of the typical user is hence
$\mathcal{R} = \sum_{k = 1}^{K} p_k \mathcal{R}^{(k)}.$

\textbf{Shannon throughput per Joule}: In $5$G NR, a key motivation for proposing the inclusion of BWP is power savings through adaptive bandwidth allocation. Hence, we consider Shannon throughput per Joule of the energy spent, which is the Shannon throughput divided by the spent power $kP$. Conditioning on the typical transmitter being of type $k$, Shannon throughput per Joule $\mathcal{R}^{(k)}_{J}$ is obtained by dividing \eqref{eq:Shannon_throughput_no_joule} by $kP$ as
\begin{align}
\mathcal{R}^{(k)}_{J} = \frac{1}{KP}\int_{0}^{\infty} p_{\rm s}^{(k)}(2^y-1)~\mathrm{d}y.
\end{align}
Here, the Shannon throughput is expressed in bits/Joule.
\begin{remark}
Note that, in this paper, the Shannon throughput per Joule is equivalent to the Shannon throughput per Hertz in that they only differ in the multiplicative constant $P$. This is due to the assumption that a transmitter spends power $P$ per chunk. Thus, for a given type of the user, normalizing by the transmit power to obtain the Shannon throughput per Joule is equivalent to normalizing by the bandwidth used by the user. 
\end{remark}

\section{The Mean Model}
\label{sec:mean_model}

\subsection{Comparison of two networks with adaptive BA}
Suppose the network operator has decided to use adaptive BA. There are several BA configurations to choose from based on the values of the probabilities $p_k$. From the operator's viewpoint, it is natural to investigate how two networks with different $p_k$ fare in terms on the overall performance. To fairly compare two networks employing adaptive BA, it is natural to do so for the same mean signal and the same mean interference powers.

\textbf{Mean signal power}: For our BA model, since the typical transmitter is of type $k$ with probability $p_k$, the mean signal power at the typical receiver is given by
\begin{equation}
\bar{S}_{k} = P\ell(x_0)\sum_{k = 1}^{K}p_kk.
\label{eq:mean_sig_bwp}
\end{equation}
For another network with different $p_k$, say $p_k'$, we might have to adjust the transmit power per chunk $P'$ to have the mean signal power $\bar{S}_{k}'$ same as $\bar{S}_{k}$. From \eqref{eq:mean_sig_bwp}, it follows that
\begin{equation}
P' = \frac{P\sum_{k = 1}^{K}k p_k}{\sum_{k = 1}^{K}k p_k'} 
\label{eq:eq:_power}
\end{equation}
leads to $\bar{S}_{k} = \bar{S}_{k}'$.

\textbf{Mean interference power}: For our BA model, conditioned on the fact that the typical user is of type $k$, the mean interference power is obtained by taking the expectation of $I_k$ given in \eqref{eq:intf_pow_main} as\vspace*{-3mm}

{{\small \begin{align}
\mathbb{E}[I_k]  &= \sum_{i = 1}^{K}\sum_{t = 0 \vee (i+k-K)}^{k \wedge i}\mathbb{E}\left[\sum_{x \in \Phi_{i,t}\setminus \lbrace x_0 \rbrace} tP\ell(x)\right] \nonumber\\
&\overset{(a)}{=} \lambda\pi \delta c_0^{\delta - 1}\Gamma(\delta)\Gamma(1-\delta)P\sum_{i = 1}^{K}p_i \sum_{t = 0 \vee (i+k-K)}^{k \wedge i} tp_{k,i}^{(t)},
\label{eq:cond_mean_intf}
\end{align}}}\vspace{-3mm}

\noindent where $(a)$ is obtained by the straightforward application of the Campbell's theorem for the PPP $\Phi_{i,t}$~\cite{bb_book} and using the bounded path loss function $\ell(x) = (c_0 +\|x\|^{\alpha})^{-1}$ with $\delta \triangleq 2/\alpha$.\footnote{The standard path loss function $\ell(x) = \|x\|^{-\alpha}$ follows from the bounded path loss function as $c_0 \to 0$.} Unconditioning on the type of the typical user yields the mean interference power at the typical receiver as
\begin{align}
\bar{I} = \sum_{k = 1}^{K} p_k \mathbb{E}[I_k].
\label{eq:mean_intf}
\end{align}

\begin{figure*}[t]
	\begin{minipage}{0.32\linewidth}
		\begin{center}
			\includegraphics[width = 2in,height=1.3in]{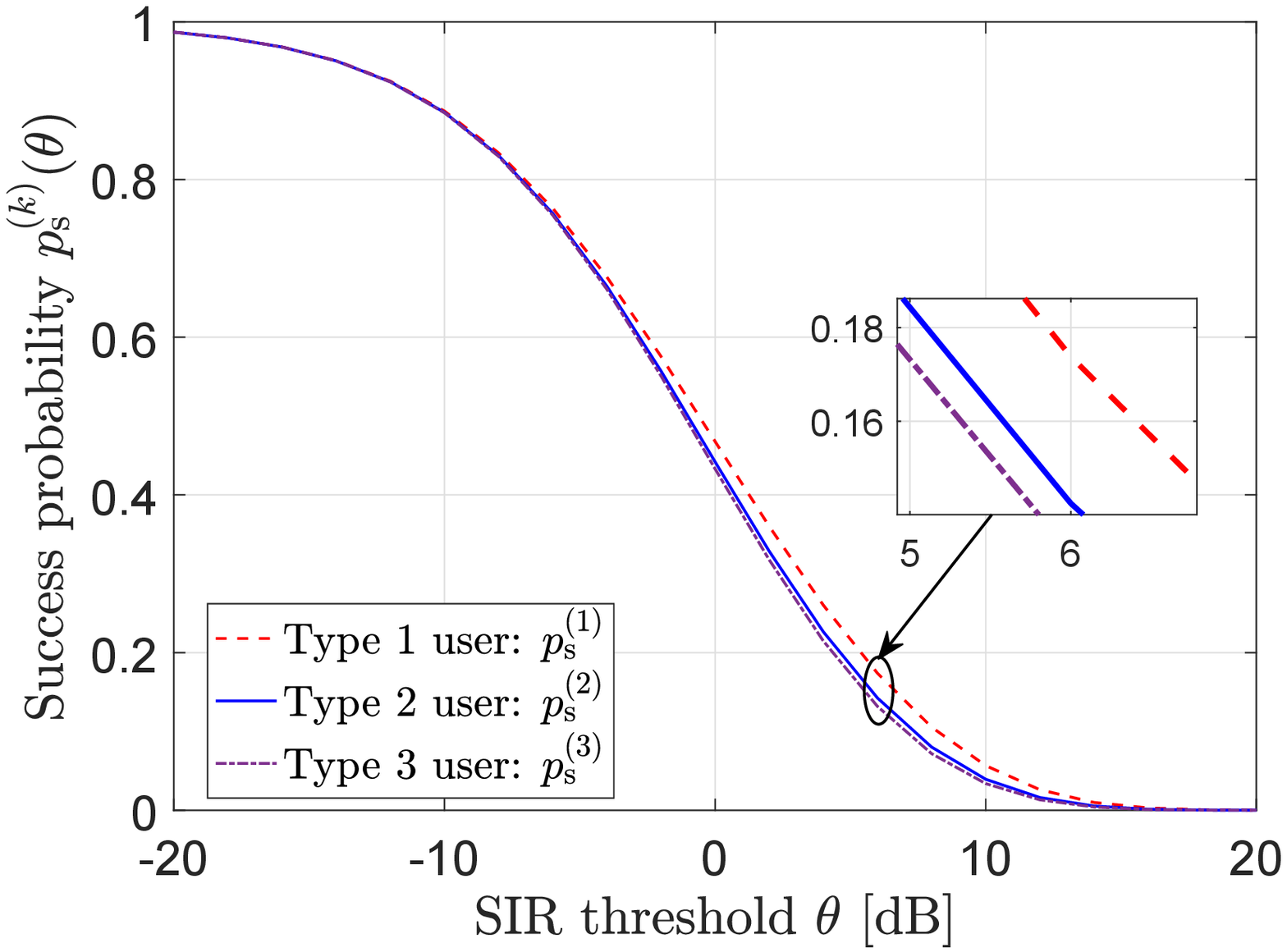}
			\caption{\footnotesize Per-type success probability $p_{\rm s}^{(k)}$ versus the SIR threshold $\theta$. $p_k = 1/K = 1/3$.}
			\label{fig:suc_prob_N_3_uniform_alpha_4}
		\end{center}
	\end{minipage}
	\hspace{0.01\linewidth}
	\begin{minipage}{0.32\linewidth}
		\begin{center}
			\includegraphics[width = 2in,height=1.3in]{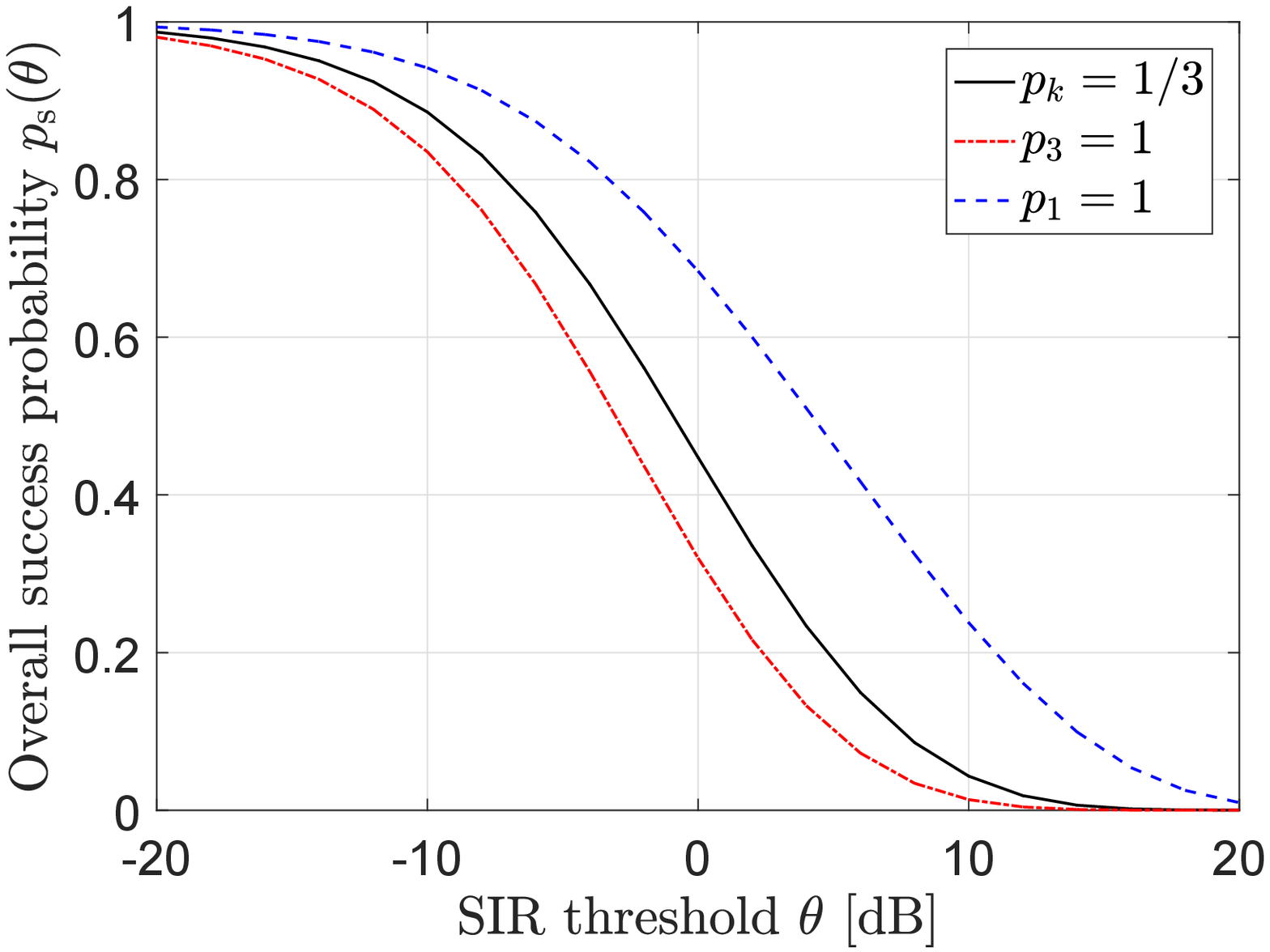}
			\caption{\footnotesize Overall success probability $p_{\rm s}$ versus the SIR threshold $\theta$.}
			\label{fig:overall_suc_prob_K_3_alpha_4}
		\end{center}
	\end{minipage}
	\hspace{0.01\linewidth}
	\begin{minipage}{0.32\linewidth}
		\begin{center}
			\includegraphics[width = 2in,height=1.3in]{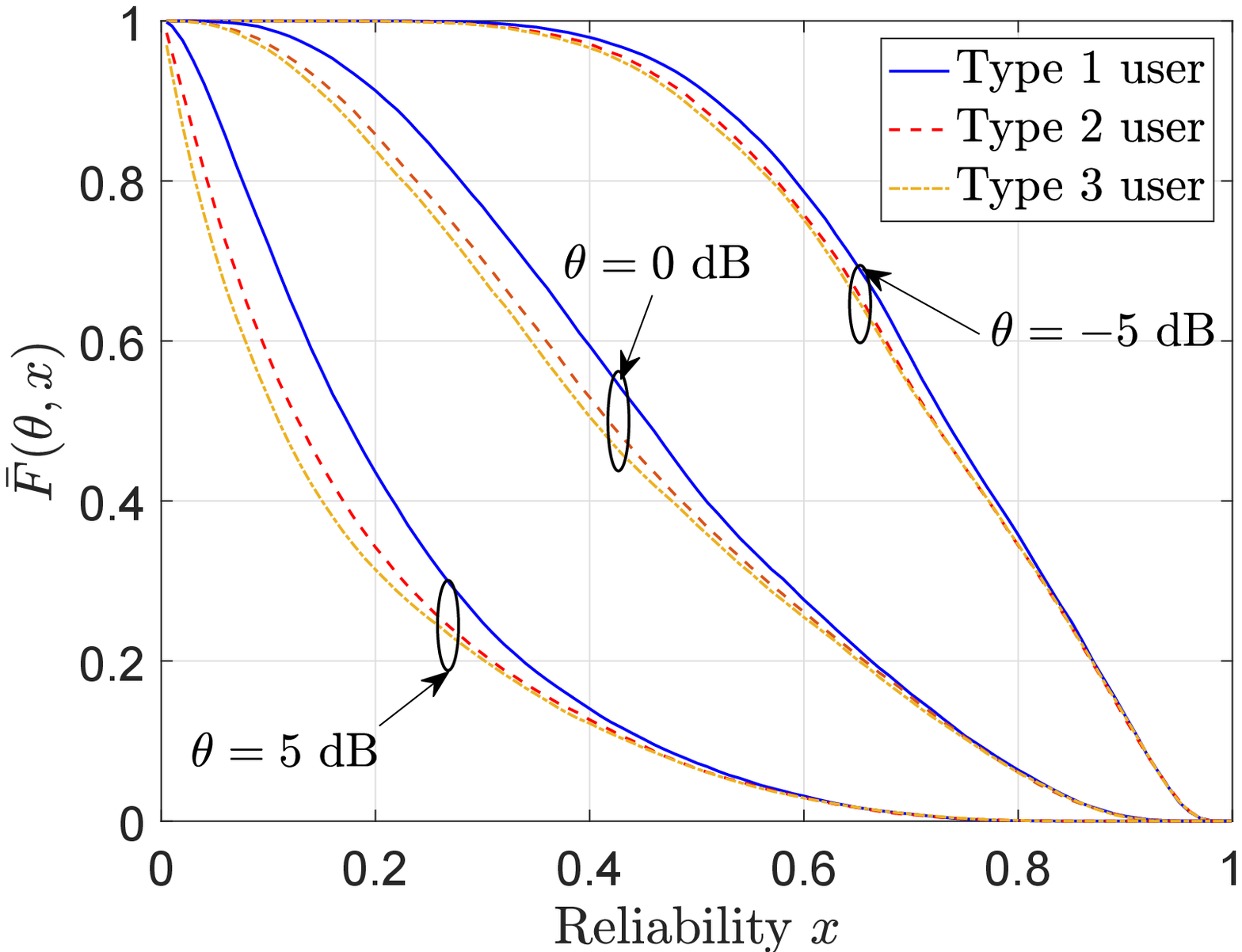}
			\caption{\footnotesize The SIR meta distribution $\bar{F}(\theta, x)$ versus the reliability threshold $x$. $p_k = 1/K = 1/3$.}
			\label{fig:md_basic}
		\end{center}
	\end{minipage}
\end{figure*}

\begin{figure*}[t]
	\begin{minipage}{0.32\linewidth}
		\begin{center}
			\includegraphics[width = 2in,height=1.3in]{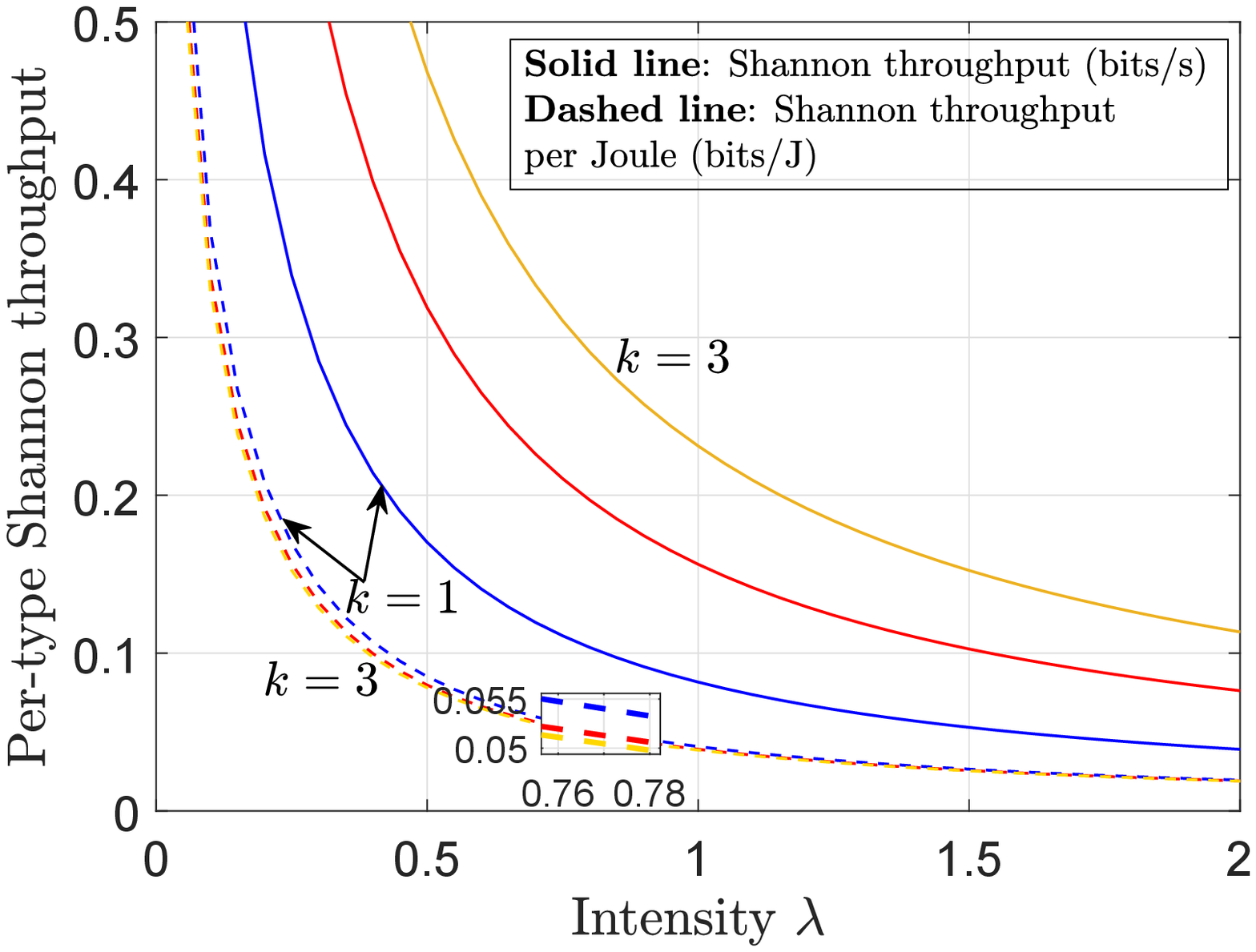}
			\caption{\footnotesize Per-type Shannon throughput versus intensity $\lambda$. $P = 2$ and $p_k = 1/K = 1/3$. For solid curves: $k = 1, 2, 3$ (from bottom to top). For dashed curves: $k = 1, 2, 3$ (from top to bottom).}
			\label{fig:shannon_rate_joule_no_joule_no_mean}
		\end{center}
	\end{minipage}
	\hspace{0.01\linewidth}
	\begin{minipage}{0.32\linewidth}
		\begin{center}
			\includegraphics[width = 2in,height=1.3in]{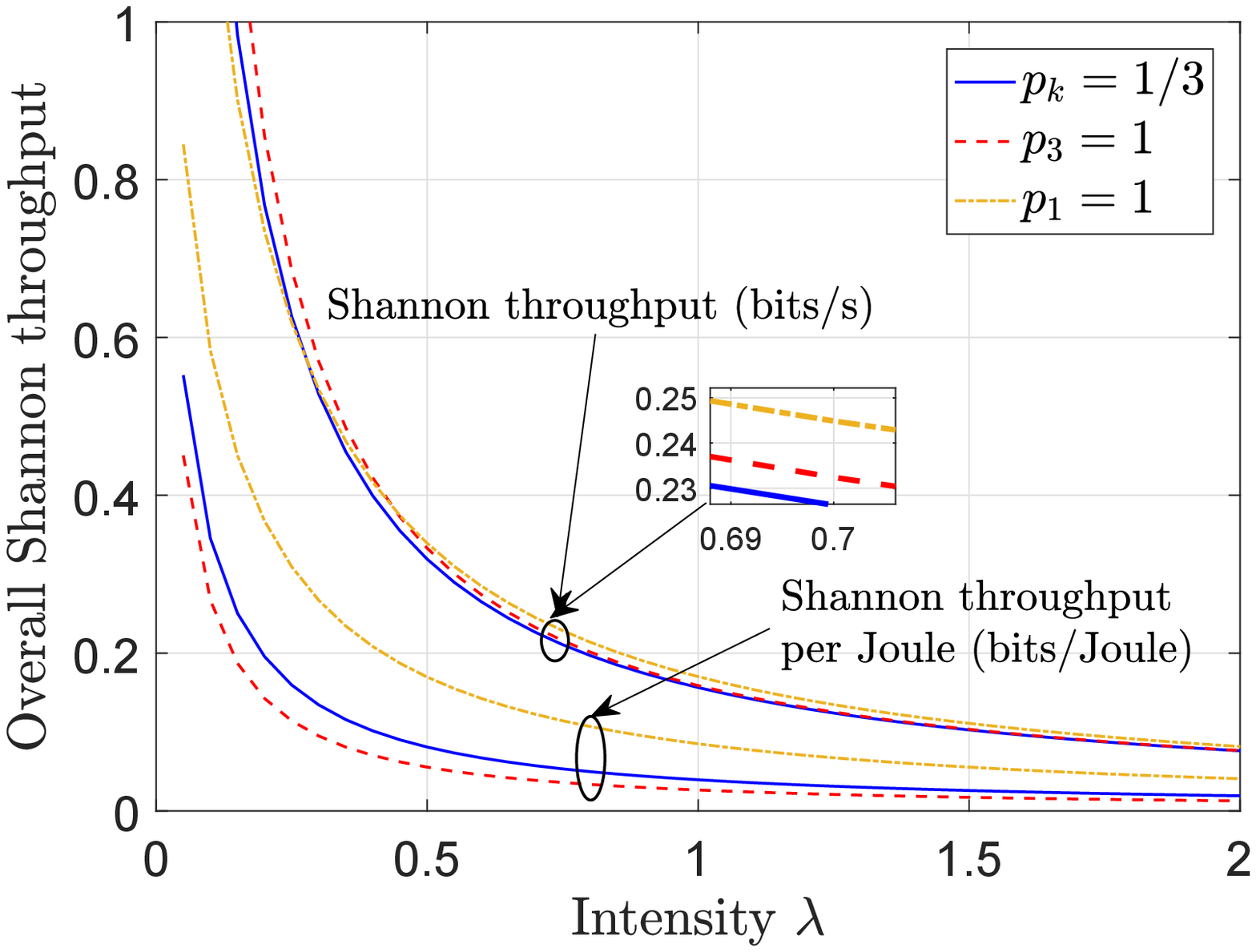}
			\caption{\footnotesize Overall Shannon throughput versus intensity $\lambda$. $P = 2$.}
			\label{fig:Overall_shannon_joule_no_joule}
		\end{center}
	\end{minipage}
	\hspace{0.01\linewidth}
	\begin{minipage}{0.32\linewidth}
		\begin{center}
			\includegraphics[width = 2in,height=1.3in]{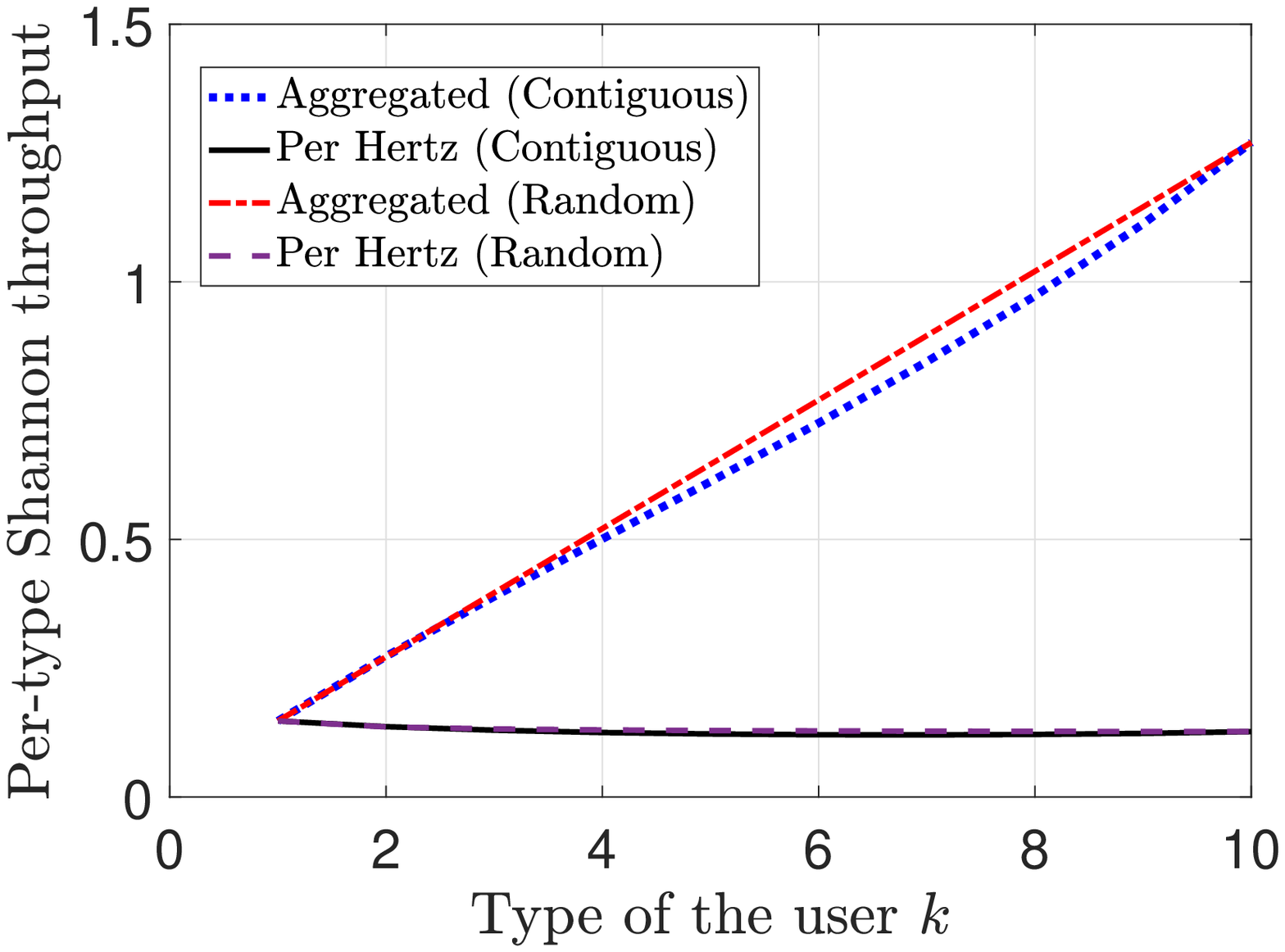}
			\caption{\footnotesize Per-type Shannon throughput versus the type $k$ of the user for random and contiguous bandwidth allocation. $K = 10$, $p_i = 1/K$.}
			\label{fig:Mean_interference_contiguous}
		\end{center}
	\end{minipage}
\end{figure*}

For another adaptive BA-based network with probabilities $p_k'$, the mean interference power $\bar{I}'$ can be calculated by replacing $p_k$ by $p_k'$ and $\lambda$ by $\lambda'$ in \eqref{eq:cond_mean_intf} and \eqref{eq:mean_intf}, where $\lambda'$ is the intensity of the PPP corresponding to the network with $p_k'$. The intensity $\lambda'$ is chosen such that the mean interference powers for two networks with $p_k$ and $p_k'$ are the same, i.e., $\bar{I} = \bar{I}'$. Hence, from \eqref{eq:cond_mean_intf} and \eqref{eq:mean_intf}, it follows that
\begin{align}
\lambda' = \frac{\lambda P\sum_{k = 1}^{K}p_k \sum_{i = 1}^{K}p_i \sum_{t = 0 \vee (i+k-K)}^{k \wedge i} tp_{k,i}^{(t)}}{P'\sum_{k = 1}^{K}p_k' \sum_{i = 1}^{K}p_i' \sum_{t = 0 \vee (i+k-K)}^{k \wedge i} tp_{k,i}^{(t)}},
\label{eq:eq_lambda}
\end{align}
which results in $\bar{I} = \bar{I}_K$.

\subsection{Service differentiation}
We now quantify the service differentiation resulting from adaptive BA. We focus on mean values of signal and interference powers to estimate the Shannon throughput (data rate) a user gets depending on its type. As shown later in the paper in Section~\ref{sec:var}, such a mean-based approach sheds light on the effect of traffic variability on the performance.
\begin{proposition}
The random BA is roughly egalitarian in Shannon throughput per Hertz and leads to a linear service differentiation in aggregate Shannon throughput.
\end{proposition}
This proposition is based on the following observation. For a user of type $k$, the mean signal power is $k$ times the mean signal power of a user of type $1$. Also, noting that \eqref{eq:pt} is the probability mass function of the hypergeometric distribution, the term  $\sum_{t = 0 \vee (i+k-K)}^{k \wedge i} tp_{k,i}^{(t)} = ik/K$ in \eqref{eq:cond_mean_intf} is the mean of the hypergeometric distribution. As a result, the mean interference at a user of type $k$ is $k$ times that at a user of type $1$. Hence, if we only think in terms of mean values, all users, regardless of their type, have the same mean signal-to-mean interference ratio (MSMIR), and hence roughly the same (egalitarian) Shannon throughput per Hertz.   Of course, since a user of type $k$ has a bandwidth $k$ times larger than that of a user of type 1, a user benefits linearly from its type in terms of aggregated Shannon throughput. This proposition is illustrated by Fig.~\ref{fig:Mean_interference_contiguous}.

Similarly, for the contiguous BA as well, as Fig.~\ref{fig:Mean_interference_contiguous} shows, the adaptive BA provides a roughly egalitarian service in Shannon throughput per Hertz and a linear service differentiation in aggregate Shannon throughput.

\section{Results and Discussions}
In this section, we discuss the results for the adaptive BA model. We divide this section in two parts. The first part provides results that focus on the per-type and the overall performance in the adaptive BA setting. The second part integrates the mean model discussed in Section~\ref{sec:mean_model}.  

Without loss of generality, we assume the following model parameters. The number of bandwidth chunks is $K = 3$ unless otherwise mentioned. The intensity of the PPP is $\lambda = 0.2$. For the bounded path loss model, the path loss exponent is $\alpha = 4$ with $c_0 = 1$. The desired link distance is $R = 1$. Other parameters are given in the captions of the relevant figures. Unless otherwise mentioned, all plots correspond to the random BA model.

\subsection{Per-type and overall performance}

\textbf{Success probability.} Fig.~\ref{fig:suc_prob_N_3_uniform_alpha_4} shows the per-type success probability. A type-$1$ user experiences a smaller interference compared to type-$2$ and type-$3$ users since it uses only one chunk selected randomly. Hence, a type-$1$ user achieves the highest success probability. As a consequence, when we uncondition on the type of the user and calculate the overall success probability, the network model where a user always selects only one chunk at random (the model with $p_1 = 1$) achieves the highest overall success probability, while a network model without adaptive BA (the case with $p_3 = 1$) performs the worst (see Fig.~\ref{fig:overall_suc_prob_K_3_alpha_4}). Other cases of adaptive BA lie in between these two extreme cases of $p_1 = 1$ and $p_3 = 1$.

\textbf{Meta distribution.} Fig.~\ref{fig:md_basic} plots the per-type SIR meta distribution against the reliability $x$ for different target SIR thresholds $\theta$. The curves in~Fig.~\ref{fig:md_basic} allow one to make precise statements about the fraction of users achieving an SIR of $\theta$ with reliability $x$. Notice that the same trend as the per-type success probability in Fig.~\ref{fig:suc_prob_N_3_uniform_alpha_4} holds for the SIR meta distribution, i.e., type-$1$ users outperform type-$2$ and type-$3$ users irrespective of the SIR threshold value. In other words, the fraction of type-$1$ users that achieve a reliability of $x$ for a given SIR threshold $\theta$ is higher than the fractions of type-$2$ and type-$3$ users. For instance, the fraction of users achieving an SIR of $-5$ dB with reliability $60\%$ is $0.78$ for type-$1$ users, $0.74$ for type-$2$ users, and $0.73$ for type-$3$ users. We skip the discussion on the overall SIR meta distribution since it follows a similar trend as that of the overall success probability in~Fig.~\ref{fig:overall_suc_prob_K_3_alpha_4}.
 
\textbf{Shannon throughput.}
For the Shannon throughput defined in~\eqref{eq:shannon_throughput}, as shown by solid curves in Fig.~\ref{fig:shannon_rate_joule_no_joule_no_mean}, a new trend emerges for the per-type Shannon throughput, where a type-$3$ user outperforms type-$2$ and type-$1$ users. This trend occurs because a higher allocated bandwidth boosts the per-type Shannon throughput, which overcomes the increased interference due to higher bandwidth. Here, note that the users of higher types achieve a higher Shannon throughput at the expense of larger transmit power $kP$ (hence more power consumption), which grows linearly with $k$ (the type of the user). 

\textbf{Shannon throughput per Joule.}\footnote{Note that, as mentioned in Remark $1$, the trends observed here in the Shannon throughput per Joule hold true for the Shannon throughput per Hertz as well.}
As shown by dashed curves in Fig.~\ref{fig:shannon_rate_joule_no_joule_no_mean}, the trend reverses when the per-type Shannon throughput is normalized by the transmit power, i.e., a type-$1$ user achieves a higher Shannon throughput than type-$2$ and type-$3$ users. This reveals the per-type throughput performance per Joule of energy spent, which is useful in understanding the effect of power consumption on the per-type throughput performance.

%


Fig.~\ref{fig:Overall_shannon_joule_no_joule} shows that the intensity $\lambda$ of the PPP plays a key role in determining the overall Shannon throughput. For small intensity $\lambda$, as shown by solid curves in Fig.~\ref{fig:Overall_shannon_joule_no_joule}, the network model with no adaptive BA (the model with $p_3 = 1$) achieves a higher overall Shannon throughput than two networks employing adaptive BA with $p_k = 1/3$ and $p_1 = 1$. In contrast, for large $\lambda$, the trend reverses in that the networks with adaptive BA outperform the network with no adaptive BA. The reason behind this behavior is that: for a small $\lambda$, the interference power is relatively small. Hence, a wide bandwidth in the network without adaptive BA boosts the Shannon throughput and overcomes the negative impact of increased interference due to higher bandwidth. Whereas for large $\lambda$, the effect of increased interference dominates and the network with adaptive BA achieves a higher Shannon throughput. 

Also, again similar to the per-type Shannon throughput case, the network without adaptive BA achieves a higher Shannon throughput for small $\lambda$ at the expense of larger transmit power $kP$ (hence more power consumption). Thus, when the Shannon throughput is normalized by the transmit power, the network with the least interference power (the network with $p_1 = 1$ in this case) achieves the highest Shannon throughput.


\begin{figure*}[t]
	\begin{minipage}{0.32\linewidth}
		\begin{center}
			\includegraphics[width = 2in,height=1.3in]{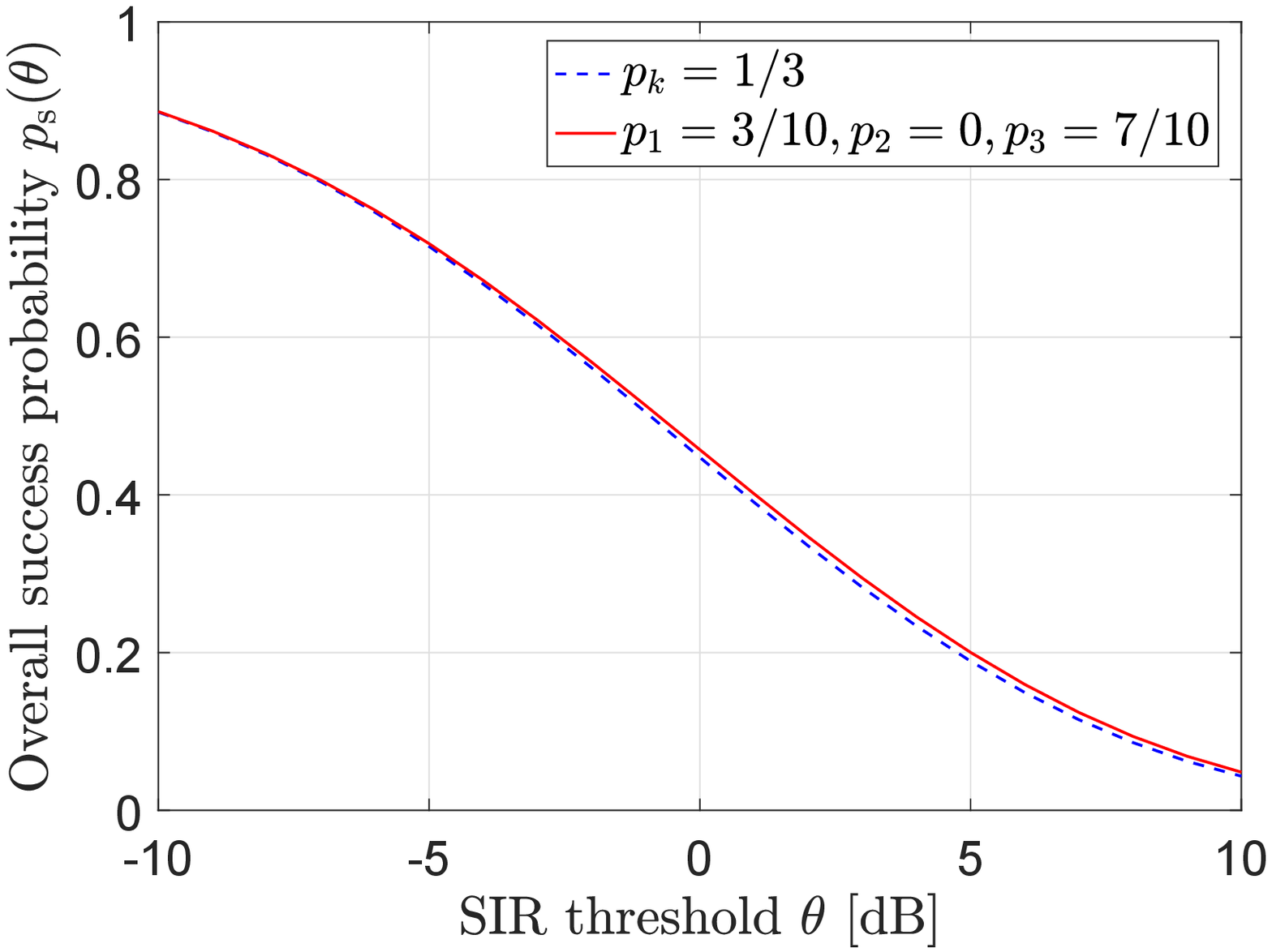}
			\caption{\footnotesize Overall success probability $p_{\rm s}(\theta)$ versus the SIR threshold $\theta$. }
			\label{fig:suc_prob_mean_model}
		\end{center}
	\end{minipage}
	\hspace{0.01\linewidth}
	\begin{minipage}{0.32\linewidth}
		\begin{center}
			\includegraphics[width = 2in,height=1.3in]{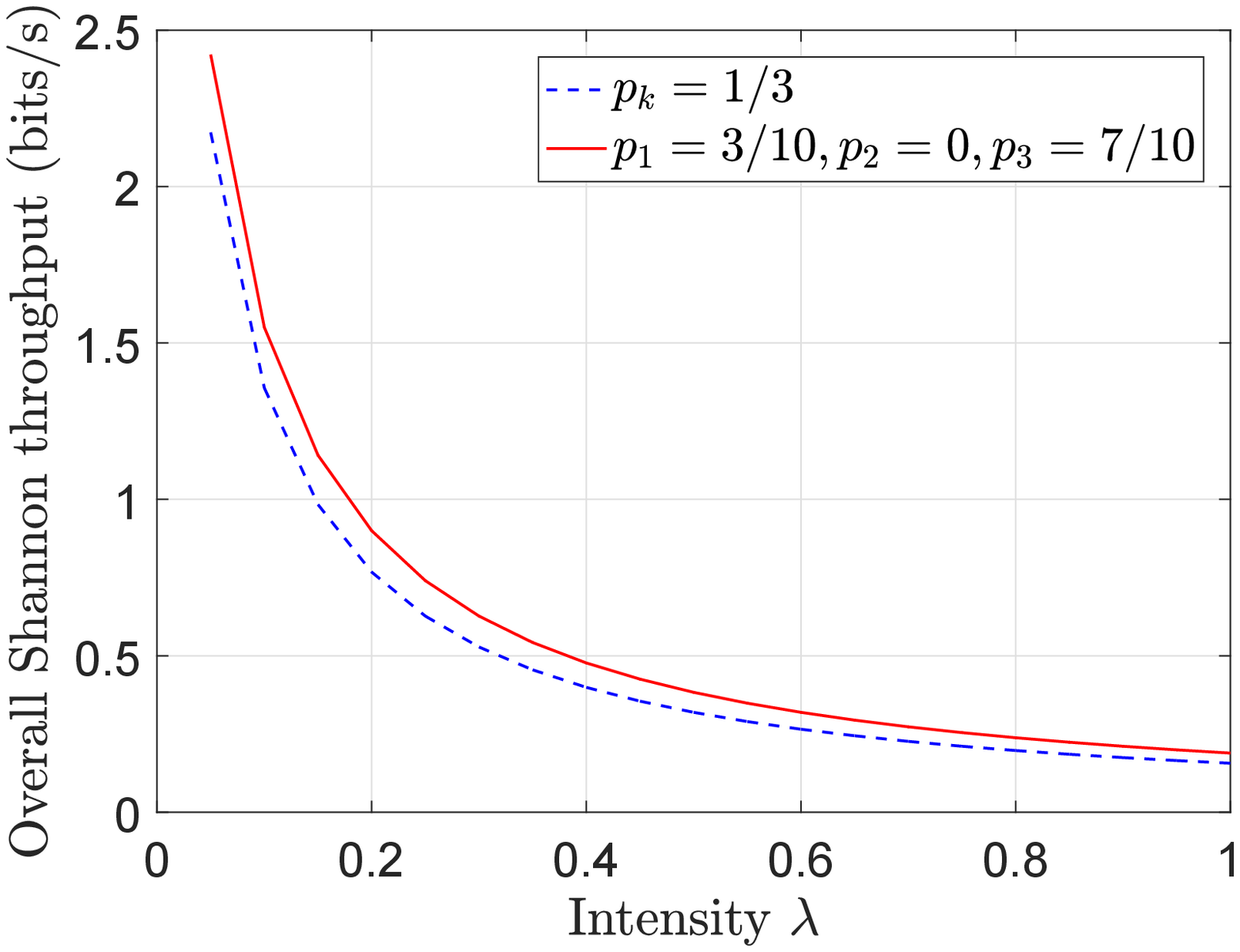}
			\caption{\footnotesize Overall Shannon throughput versus intensity $\lambda$. $P = 2$. }
			\label{fig:shannon_rate_3_10_7_10_N_3_mean_model}
		\end{center}
	\end{minipage}
	\hspace{0.01\linewidth}
	\begin{minipage}{0.32\linewidth}
		\begin{center}
			\includegraphics[width = 2in,height=1.3in]{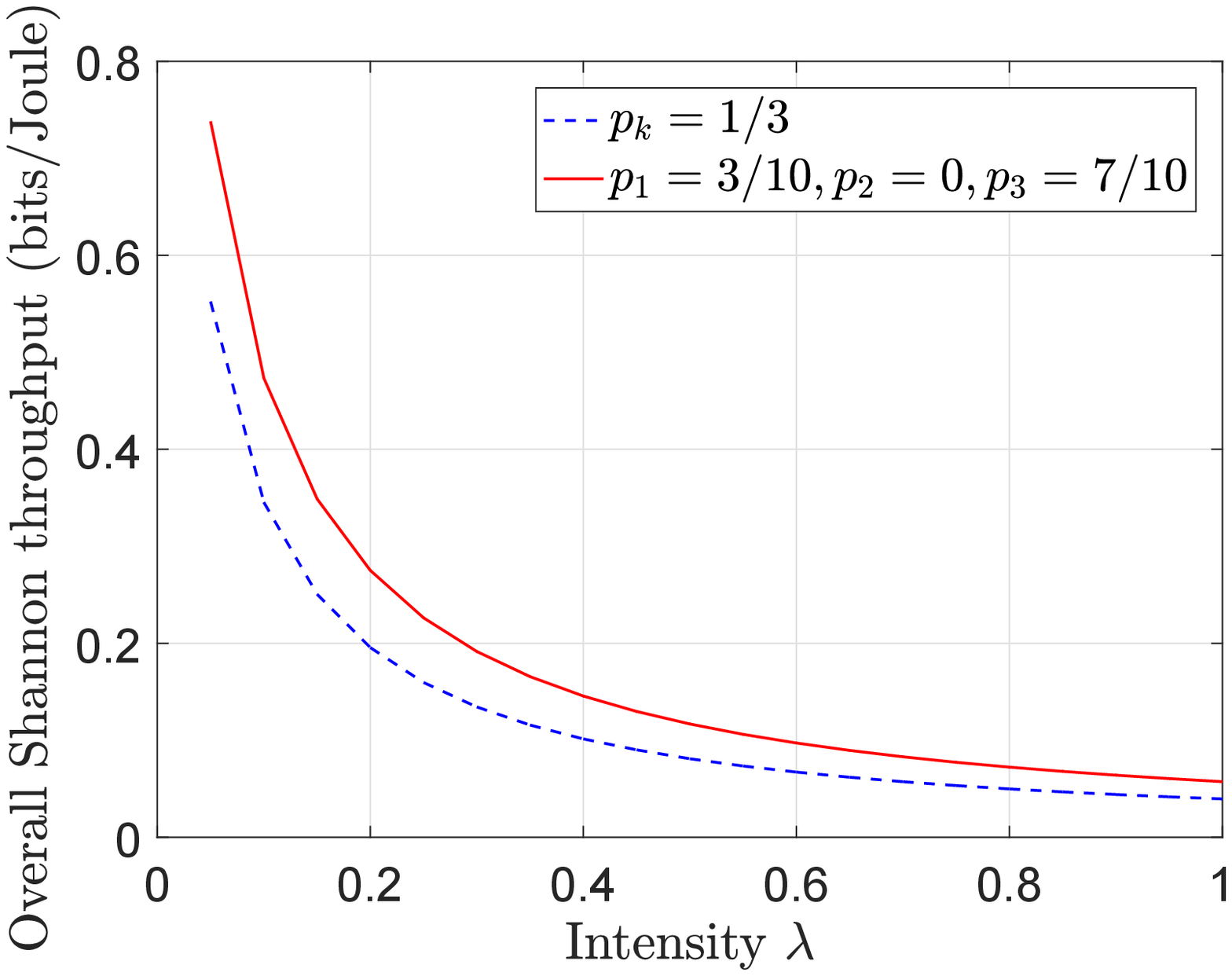}
			\caption{\footnotesize Overall Shannon throughput per Joule versus intensity $\lambda$. $P = 2$.}
			\label{fig:shannon_rate_3_10_7_10_N_3_Joule_mean_model}
		\end{center}
	\end{minipage}
\end{figure*}
\textbf{Service differentiation.} Fig.~\ref{fig:Mean_interference_contiguous} plots the per-type aggregated Shannon throughput and per-Hertz Shannon throughput against the user type $k$. The plots confirm Proposition $1$ that both random and contiguous BA result in a roughly equal Shannon throughput per Hertz to users of different types and a linear increase in aggregated throughput with user type. As shown by the dashed curves in Fig.~\ref{fig:shannon_rate_joule_no_joule_no_mean}, this egalitarian property holds for different intensities $\lambda$ of the PPP. Also, as $k$ increases, the Shannon throughput per Hertz decreases slightly. This can be attributed to the fact that a type-$k$ user experiences interference with a larger variance than a type-$j$ user if $k< j$, and such a higher variance is generally beneficial to most metrics as shown in~\cite{Lee_F}. 
Another interesting observation from Fig.~\ref{fig:Mean_interference_contiguous} is that random BA results in a better Shannon throughput compared to contiguous BA since the former leads to higher traffic variability that benefits users in terms of data rates. We discuss this observation in more detail in Section~\ref{sec:var}.

\subsection{Benefits of adaptive BA}

We now briefly discuss the benefits of adaptive BA from these numerical results. Let us start with the user performance viewpoint. It should be clear from Figs.~\ref{fig:suc_prob_N_3_uniform_alpha_4} and \ref{fig:md_basic} that smaller users (i.e., users of type 1) get a better success probability and a better meta distribution than bigger ones. Note that the improvement is more pronounced for higher SIR thresholds. Fig.~\ref{fig:shannon_rate_joule_no_joule_no_mean} shows that the Shannon throughput per Joule is also better for smaller users than for
bigger ones. We conclude that adaptive BA brings the expected service differentiation and protection of small users, both in terms of success probability and Shannon throughput per Joule. As for bigger users, we see in Fig.~\ref{fig:shannon_rate_joule_no_joule_no_mean} that they nevertheless get a better Shannon throughput than smaller ones, to the
expense of a higher power consumption (proportional to the number of chunks they use). One also gets from first principles that the biggest users get less interference and hence a better Shannon throughput in the scenario with adaptive BA than in the scenario without adaptive BA. Hence, at least for this performance metric, adaptive BA is beneficial to bigger users as well.

Consider now the point of view of operators, namely overall performance, which can be linked to revenue. When comparing the uniform case to the case with $p_3 = 1$ (no adaptive BA case) in Fig.~\ref{fig:overall_suc_prob_K_3_alpha_4}, we see that the overall success probability is higher in the situation with adaptive BA than in the situation without.
Fig.~\ref{fig:Overall_shannon_joule_no_joule} actually shows that the same conclusion holds for the overall Shannon throughput per Joule. We conclude that adaptive BA should be beneficial to operators as well. These conclusions are not limited to this special case with three chunks as the ordering of the curves is in fact the same when varying the number of chunks and/or the other model parameters.

\subsection{Increasing traffic variability may improve performance}
\label{sec:var}
Suppose the operator has decided to use adaptive BA and is interested in comparing the overall performance of two networks with different probabilities $p_k$. Consider a first network with uniformly distributed $p_k$, for $K = 3$, $p_k = 1/K = 1/3$. For the second network, also with $K = 3$, assume that $p_3 = 0.7$ and $p_1 = 0.3$. The choice of $p_k$ for the second network is inspired from the estimated proportion of the video traffic for year 2020~\cite{Cisco_VNI}. It is expected that approximately $70\%$ of mobile traffic will correspond to videos needing a wide bandwidth. Hence, $p_3 = 0.7$ corresponds to video traffic, while $p_1 = 0.3$ corresponds to non-video traffic due to applications requiring smaller bandwidths. For the second network, the transmit power per chunk $P'$ and the intensity $\lambda'$ are calculated from \eqref{eq:eq:_power} and \eqref{eq:eq_lambda}, respectively, such that both networks have the same mean interference and the same mean signal powers.

Compared to the first network, the second one has a more variable distribution of powers and hence a more variable interference. From Figs.~\ref{fig:suc_prob_mean_model}, \ref{fig:shannon_rate_3_10_7_10_N_3_mean_model}, and \ref{fig:shannon_rate_3_10_7_10_N_3_Joule_mean_model}, it is clear that the second and more variable traffic network outperforms the first one in terms of all performance metrics: success probability, Shannon throughput, and Shannon throughput per Joule. 

For the Shannon throughput case, the trends in both the Shannon throughput and the Shannon throughput per Joule with intensity $\lambda$ are the same (see Figs.~\ref{fig:shannon_rate_3_10_7_10_N_3_mean_model} and \ref{fig:shannon_rate_3_10_7_10_N_3_Joule_mean_model}). This is different from the case without equality of the mean values as shown in Fig.~\ref{fig:Overall_shannon_joule_no_joule}, where we observed that, for small $\lambda$, the trend in Shannon throughput per Joule is opposite of that in Shannon throughput. Also, there is no a crossover in curves of Shannon throughput with $\lambda$ in the model with balanced means.

%
%
%
%

\section{Concluding Remarks and Future Directions}
This paper proposes a first analytic model for the prediction of the adaptive BA inspired by BWP. The proposed model allowed us to show that adaptive BA should benefit to both small and big users.
We showed that small users are well protected by adaptive BA in terms of success probability, meta distribution of the SIR, and Shannon throughput per Joule. On the other hand, big users achieve a better Shannon throughput than in the situation without adaptive BA. The analysis of overall performance allowed us to show that adaptive BA should also be beneficial to operators. Also, we observe that adaptive BA is roughly egalitarian per Hertz and leads to a linear service differentiation in aggregated Shannon throughput. 

There are several future directions of research. A natural extension is to study adaptive BA in cellular settings. It would also be interesting to see how to strategically assign bandwidth chunks to users based on their local environment. Finally, this work is limited to a snapshot analysis of the network. The inclusion of dynamics will certainly be of great interest as well.

\section*{Acknowledgment}
The authors would like to thank Luis Guilherme Uzeda Garcia, Fuad Abinader, Andrea Marcano, and Dalia Popescu from Nokia Bell Labs, Nozay, France for the initial discussion on this problem and bringing to our attention the lack of analytical formulation of BWP.

\appendices
\section{Proof of Lemma~\ref{lem:suc_prob_k}}
\label{app:suc_prob_k}
Conditioned on $x_0 \in \Phi_k$, the success probability $p_{\rm s}^{(k)}$ is\vspace*{-3mm}

{{\small \begin{align*}
&p_{\rm s}^{(k)}(\theta)  = \mathbb{P}(\mathsf{SIR}_{o}^{(k)} > \theta \mid x_0 \in \Phi_k) \nonumber \\
&= \mathbb{P}\!\left(\!h_{x_0} > \theta \frac{\sum_{i = 1}^{K} \sum_{t = 0 \vee (i+k-K)}^{k \wedge i}\sum_{x \in \Phi_{\Phi_{i,t}}\setminus \lbrace x_0 \rbrace}th_x \ell(x)}{k\ell(x_0)}\!\right) \nonumber \\
&= \mathbb{E}\!\left[\!\exp\!\left(\!\!-\frac{\theta \sum_{i = 1}^{K} \sum_{t = 0 \vee (i+k-K)}^{k \wedge i}\sum_{x \in \Phi_{\Phi_{i,t}}\setminus \lbrace x_0 \rbrace}t h_x\ell(x)}{k\ell(x_0)}\!\right)\!\!\right]\!\!.
\end{align*}}}
Averaging over interfering fading channels, it follows that
\begin{equation}
p_{\rm s}^{(k)}(\theta)  =\prod_{i=1}^{K} \underbrace{\prod_{t = 0 \vee (i+k-K)}^{k \wedge i} \mathbb{E}\left[\prod_{x \in \Phi_{i,t} \setminus \lbrace x_0 \rbrace} \frac{1}{1+\frac{\theta t \ell(x)}{k\ell(x_0)}}\right]}_{p_{{\rm s} \mid k}^{(i)}(\theta)}.
\label{eq:suc_ki_inter}
\end{equation}
The probability $p_{\rm s}^{(k,i)}$ can be interpreted as the success probability of the typical receiver due to interference from type-$i$ interferers only conditioned on the typical transmitter being of type $k$. The reason behind this interpretation is that the expression of $p_{\rm s}^{(k,i)}$ in \eqref{eq:suc_ki_inter} can be obtained by calculating $
\mathbb{P}\left(\frac{S}{I_{k,i}} > \theta\right) = \mathbb{P}\left(\mathsf{SIR}_o^{(k, i)} > \theta\right)$, where $S$ is signal power and $I_{k,i}$ given by \eqref{eq:intf_pow_i} is the interference power received at the typical receiver only from type-$i$ interferers. Thus, $\mathsf{SIR}_o^{(k,i)}$ is the SIR at the typical receiver when the interference from type-$i$ interferers only is taken into account.

\section{Proof of Theorem~\ref{thm:suc_ki}}
\label{app:suc_ki}
Continuing from \eqref{eq:suc_ki_inter}, we have
\begin{align}
p_{\rm s}^{(k,i)}(\theta) = \prod_{t = 0 \vee (i+k-K)}^{k \wedge i} \mathbb{E}\left[\prod_{x \in \Phi_{i,t} \setminus \lbrace x_0 \rbrace} \frac{1}{1+\frac{\theta t \ell(x)}{k\ell(x_0)}}\right].
\end{align}
For the power-law path loss model $\ell(r) = r^{-\alpha}$, by the probability generating functional (PGFL) of the PPP, it follows that
\begin{align}
p_{\rm s}^{(k,i)}(\theta) = \hspace*{-6mm}\prod_{t =0 \vee (i+k-K)}^{k \wedge i}\hspace*{-6mm}\exp\left(-\lambda_{i,t}\int_{\mathbb{R}^2}\left(1- \frac{1}{1+\frac{t\theta R^{\alpha} \|x\|^{-\alpha}}{k}}\right)\mathrm{d}x\right)\!\!,
\label{eq:suc_ki_int}
\end{align}
where $\lambda_{i,t}$ is the intensity of the point process $\Phi_{i,t}$. Here, $\lambda_{i,t} = \lambda_i p_{k,i}^{(t)}$ since the PPP $\Phi_i$ of interferers of type $i$ is partitioned into $t$ independent PPPs of interferers of type $i$ having $t$ common chunks with the typical transmitter. Solving the integral in \eqref{eq:suc_ki_int} and substituting $\lambda_i = \lambda p_i$, we have the desired expression of $p_{\rm s}^{(k,i)}$.

\section{Proof of Theorem~\ref{thm:Mb_k}}
\label{app:Mb_k}
Conditioning on the typical transmitter $x_0$ being of type $k$, the conditional success probability $P_{\rm s}^{(k)}$ is given as
\begin{align*}
P_{\rm s}^{(k)}(\theta) &= \mathbb{P}(\mathsf{SIR}_o^{(k)} > \theta \mid \Phi) \nonumber \\
&=  \mathbb{P}\left(h_{x_0} > \theta \frac{\sum_{x \in \Phi\setminus\lbrace x_0 \rbrace} t_x h_x  \ell(x)}{k\ell(x_0)} \mid \Phi\right)  \\ 
& = \prod_{x \in \Phi\setminus\lbrace x_0\rbrace} \mathbb{E}\left[\exp\left(-\theta \frac{ t_x h_x  \ell(x)}{k\ell(x_0)}\right) \mid  \Phi\right],
\end{align*}
where the expectation is taken over the random channel access scheme of interferers determined by the BWP model and the fading. Recall that, for the BWP model, each interferer can be of type $i$ with probability $p_i$. Then, a type-$i$ interferer can have $ 0 \vee (i+k-K)\leq t \leq k \wedge i$ chunks common with the typical transmitter with probability $p^{(t)}_{k,i}$. Thus, by averaging over the channel access scheme, it follows that\vspace*{-3mm} 

{{\small \begin{align*}
P_{\rm s}^{(k)}(\theta) = \hspace*{-3mm}\prod_{x \in \Phi  \setminus \lbrace x_0 \rbrace}\!\!\!\mathbb{E}\!\left[\sum_{i = 1}^{K}p_i\hspace*{-3mm}\sum_{t = 0 \vee (i+k-K)}^{k \wedge i}\hspace*{-3mm}p^{(t)}_{k,i}\exp\!\left(-\theta \frac{ t h_x  \ell(x)}{k\ell(x_0)}\right)\mid \Phi\right]\!\!.
\end{align*}}}
Now, averaging over the fading on interfering channels yields
\begin{align*}
P_{\rm s}^{(k)}(\theta) =  \prod_{x \in \Phi  \setminus \lbrace x_0 \rbrace}\sum_{i = 1}^{K}p_i\sum_{t = 0 \vee (i+k-K)}^{k \wedge i}\frac{p^{(t)}_{k,i}}{1+\theta \frac{t}{k}\frac{\ell(x)}{\ell(x_0)}}.
\end{align*}
The $b$th ($b \in \mathbb{C}$) moment of $P_{\rm s}^{(k)}$ can be expressed as
{{\small\begin{align*}
M_{b}^{(k)} = \exp\!\!\left(\!\!-2\pi\lambda \!\!\int_{0}^{\infty}\!\!\left[\!1-\!\left(\sum_{i = 1}^{K}p_i\hspace*{-3mm}\sum_{t = 0 \vee (i+k-K)}^{k \wedge i}\hspace*{-1mm}\frac{p^{(t)}_{k,i}}{1+\theta \frac{t}{k}\frac{\ell(r)}{\ell(x_0)}}\!\!\right)^{b}\right]\!\mathrm{d}r\!\!\right)\!\!,
\end{align*}}}
when making use of the PGFL of the PPP.

\end{document}